\documentclass[aps,amsmath, amssymb, 
superscriptaddress, 
twocolumn, 
pra, 
showpacs,
showkeys,
floatfix
]{revtex4-2}
\usepackage{graphicx}
\usepackage{color}
\usepackage{hyperref}
\usepackage{CJK}

\begin{document}
\begin{CJK*}{UTF8}{gbsn}

\title{General mapping of one-dimensional non-Hermitian mosaic models to non-mosaic counterparts: Mobility edges and Lyapunov exponents}
\author{Sheng-Lian Jiang (蒋盛莲)}
\affiliation{School of Physics, South China Normal University, Guangzhou 510006, China}
\author{Yanxia Liu (刘彦霞)}
\email{yxliu-china@ynu.edu.cn}
\affiliation{School of Physics and Astronomy, Yunnan University, Kunming 650091, China}
\author{Li-Jun Lang (郎利君)}
\email{ljlang@scnu.edu.cn}
\affiliation{School of Physics, South China Normal University, Guangzhou 510006, China}
\affiliation{Guangdong Provincial Key Laboratory of Quantum Engineering and Quantum Materials, School of Physics, South China Normal University, Guangzhou 510006, China}

\date{\today}

\begin{abstract}
	We establish a general mapping from one-dimensional {non-Hermitian} mosaic models to their non-mosaic counterparts. This mapping can give rise to mobility edges and even Lyapunov exponents in the mosaic models if critical points of localization or Lyapunov exponents of localized states in the corresponding non-mosaic models have already been analytically solved. 
	To demonstrate the validity of this mapping, we apply it to two {non-Hermitian} localization models: an Aubry-Andr\'{e}-like model with nonreciprocal hopping and complex quasiperiodic potentials, and the Ganeshan-Pixley-Das Sarma model with nonreciprocal hopping. We successfully obtain the mobility edges and Lyapunov exponents in their mosaic models.
	This general mapping may catalyze further studies on mobility edges, Lyapunov exponents, and other significant quantities pertaining to localization in {non-Hermitian} mosaic models.
\end{abstract}

	

\maketitle
\end{CJK*}

\section{Introduction}

The effective non-Hermitian Hamiltonians can be used to describe open quantum systems \cite{BreuerPetruccione2002}. Non-Hermiticity can cause intriguing properties, such as exceptional points,  non-Hermitian skin effects, non-Hermitian topological phenomena, and the non-Bloch band theory \cite{Xu-Duan-2017,GongUeda2018,KunstBergholtz2018,YaoWang2018,LeeThomale2019,KawabataSato2019,XiaoXue2019,HelbigThomale2020,BorgniaSlager2020,OkumaSato2020,WeidemannSzameit2020,Longhi2020,BergholtzKunst2021,WangFan2021,WangXue2021}, which have no counterparts in Hermitian systems. Meanwhile, non-Hermiticity in disorder or quasiperiodic systems also brings up new phenomena, for example, the mobility edges can emerge separating localized and extended states in the complex plane of spectrum in the Hatano-Nelson model, a prototypical one-dimensional (1D) non-Hermitian model characterized by nonreciprocal hopping with random on-site disorder \cite{HatanoNelson1996,HatanoNelson1998}. It is well known that an arbitrarily small strength of random disorder leads to the localization of all eigenstates in one- and two-dimensional Hermitian models \cite{AbrahamsRamakrishnan1979,LeeRamakrishnan1985,EversMirlin2008}, and the effects of disorders in corresponding non-Hermitian models have also been studied recently \cite{TzortzakakisEconomou2020,HuangShklovskii2020}. Besides the non-Hermitian models with random disorders, the non-Hermitian quasiperiodic systems have also sparked a great deal of interest \cite{JazaeriSatija2001,Yuce2014,ZengLu2017,JiangChen2019,Longhi2019a,Longhi2019b,ZengXu2020a,LiuChen2020,ZengXu2020b,LiuLonghi2020,LiuChen2021a,LiuChen2021b,WangLiu2021,Cai2021,GongCheng2021,LiuChen2021c,DwiputraZen2022}, especially the non-Hermitian variants \cite{JiangChen2019,Longhi2019a,Longhi2019b} of the celebrated Aubry-Andr\'{e} model \cite{AubryAndre1980}.

Recently, Anderson localization and mobility edges have been regaining much attention in quasiperiodic systems \cite{Biddle2010,Deng2019,Yao2019,WangLiu2020,Roy2021,LiuPalencia2022}, because the critical points of localization in these models can be analytically solved with the aid of self-duality \cite{AubryAndre1980} or further Avila's global theory \cite{Avila2015}. 
Another important reason is that quasicrystals can be experimentally realized for both Hermitian and non-Hermitian versions, which promotes the studies of localization physics and mobility edges \cite{LuschenBloch2018,AnGadway2021,WangJia2022,LinXue2022}.  The mobility edge is one of the central concepts in condensed matter physics and can be obtained in many Hermitian and non-Hermitian disorder or quasiperiodic systems \cite{Luo2021,Goblot2020,Roushan2017,Lahini2009}. The search for new typical models that can be exactly solved with mobility edges is still ongoing case by case.

In this work, without arduously dealing with models case by case, we establish a general mapping of 1D non-Hermitian mosaic models to their non-mosaic counterparts of which the critical points of localization or even the Lyapunov exponents (LEs) of localized states have been analytically solved. 
By the mapping, we can obtain the mobility edges as well as the LEs of the mosaic models. 
For the purpose of demonstration, we take two examples of {non-Hermitian} non-mosaic models to explore the localization properties of their mosaic counterparts. One is an Aubry-Andr\'{e}-like (AA-like) model with nonreciprocal hopping and complex quasiperiodic potentials \cite{LiuChen2021c}. We reconstruct the mobility edges previously obtained by Avila's global theory \cite{Avila2015}, and further derive the LEs. 
The second example involves the Ganeshan-Pixley-Das Sarma (GPD) model \cite{GaneshanDasSarma2015} with nonreciprocal hopping. Using the general mapping, we successfully obtain the analytical expressions for the mobility edges and the LEs of its mosaic counterpart, which we validate through numerical calculations.

This general mapping elucidates the mechanism behind the emergence of mobility edges in non-Hermitian mosaic models, as well as the asymmetric localization induced by nonreciprocal hopping, which can be characterized by two LEs. Furthermore, this work extends the mapping in one and two dimensions \cite{Liu2022,WangWang2022} to the {non-Hermitian} regime.


The rest of the paper is organized as follows. 
In Sec. \ref{sec:model}, we introduce a class of 1D {non-Hermitian} mosaic models. Then, we analytically derive a general mapping from this model to its non-mosaic counterpart in Sec. \ref{sec:map}. 
Subsequently, in Sec. \ref{sec:app}, we apply this mapping to two specific {non-Hermitian} mosaic models and obtain the mobility edges and the LEs.  
Finally, we summarize the results and provide a conclusion with some discussions in Sec. \ref{sec:conclusion}.

\section{The non-Hermitian mosaic models}\label{sec:model}
We consider a general class of 1D {non-Hermitian} mosaic models, which can be described by the Hamiltonian
\begin{eqnarray}
\hat{H}=\sum_j{\left( J_L|j\rangle  \langle j+1| +J_R \right|j+1 \rangle \langle j| +V_j |j  \rangle \langle j| )},
\label{eq:Ham}
\end{eqnarray}
where {$J_{R,L}\in\mathbb{C}$} denote the {complex} strengths of the nonreciprocal hopping between nearest-neighbor sites with the subscripts representing the hopping directions, and {$V_j\in\mathbb{C}$} is the {complex} on-site potential at site $j$ of the form
\begin{eqnarray}
V_j=\left\{ 
\begin{array}{cl}
	\lambda \Delta _j, & j=\kappa m ~(m\in\mathbb{Z}) \\
	0,& \mathrm{otherwise}\\
\end{array} 
\right.,
\label{eq:Disord} 
\end{eqnarray}
with {$\lambda\in\mathbb{C}$} being the {complex} strength of the mosaic potential, $\kappa \ge 1$ being an integer representing the period of non-mosaic sites $j=\kappa m$, and {$\Delta_j\in\mathbb{C}$} being a model-dependent function that can induce the localization (e.g., randomly distributed functions, quasi-periodic functions, linear functions, etc.). Every successive $\kappa$ sites form a so-called quasicell labeled by $m$, which also denotes the $m$th non-mosaic site that has a nonzero on-site potential.
For convenience, in the following we take $N$ quasicells, i.e., $m=1,\cdots,N$, and thus the system size $L=\kappa N$. 

By taking $|\Psi \rangle =\sum_j{\psi _j}|j\rangle$, the static Schr\"{o}dinger equation $\hat{H}|\Psi \rangle=E|\Psi \rangle$ can be written in terms of the amplitude $\psi_j$ as
\begin{eqnarray}
	\left\{
\begin{array}{rcl}
		E\psi _{\kappa m}&=&J_R\psi _{\kappa m-1}+J_L\psi _{\kappa m+1}+\lambda \Delta_{\kappa m}\psi _{\kappa m} \\
		E\psi _{j}&=&J_R\psi _{j-1}+J_L\psi _{j+1},~~(j\ne m\kappa) 
\end{array}
	\right..
	\label{eq:recursion}
\end{eqnarray}	 
{Note that due to the non-Hermiticity of the Hamiltonian $\hat{H}$, the eigenenergy $E$ can generally be a complex value.}

\section{Non-Hermitian mosaic-to-non-mosaic mapping}\label{sec:map}

Since there are $(\kappa-1)$ sites that are free of on-site potentials between two nearest-neighbor non-mosaic sites, a reasonable inference is that the localization, if occurs, should exist only at the non-mosaic sites. Therefore, one may be curious about the effective coupling between the non-mosaic sites. To this end, we can resort to the transfer matrix. 

From the second line of Eq. \eqref{eq:recursion}, the system of equations for the mosaic sites between two nearest non-mosaic sites, say the $m$th and the $(m+1)$th non-mosaic sites, can be given explicitly as follows
\begin{eqnarray}
	\left\{
	\begin{array}{rcl}
		E\psi _{\kappa m +1}&=&J_R\psi _{\kappa m}+J_L\psi _{\kappa m +2}\\
	 &&\cdots\\
	E\psi _{\kappa (m+1)-1}&=&J_R\psi _{\kappa (m+1)-2}+J_L\psi _{\kappa (m+1)}		
	\end{array}
	\right..
	\label{eq:Equation}
\end{eqnarray}	

With the aid of the forward transfer matrix $F$, which is defined as
\begin{eqnarray}
	\left( \begin{matrix}
		\psi_{\kappa m+2}\\
		\psi_{\kappa m+1}\\
	\end{matrix} \right)
	&=&\left( 
	\begin{matrix}
		E/J_L&		-J_R/J_L\\
		1&		0\\
	\end{matrix} \right)
	\left( \begin{matrix}
	\psi_{\kappa m+1}\\
	\psi_{\kappa m}\\
\end{matrix} \right) \notag\\
&\equiv&F\left( 
\begin{matrix}
	\psi_{\kappa m+1}\\
	\psi_{\kappa m}\\
\end{matrix} \right),
	\label{eq:Transfer}
\end{eqnarray}
Eq. \eqref{eq:Equation} can be written as
\begin{eqnarray}
\left( \begin{matrix}
	\psi_{\kappa (m+1)}\\
	\psi_{\kappa (m+1)-1}\\
\end{matrix} \right)
=F^{\kappa -1}
\left( \begin{matrix}
	\psi_{\kappa m+1}\\
	\psi_{\kappa m}\\
\end{matrix} \right),
\end{eqnarray}
and thus, the amplitude at mosaic site $\psi_{\kappa m+1}$ can be expressed in terms of the amplitudes, $\psi_{\kappa m}$ and $\psi_{\kappa (m+1)}$, of two non-mosaic sites, yielding
\begin{eqnarray}
	\psi _{\kappa m+1}=\frac{J_{L}^{\kappa -1}}{A_\kappa}\psi _{\kappa (m+1)}+\frac{J_RA_{\kappa-1}}{A_{\kappa}}\psi _{\kappa m},
	\label{eq:wave function2}
\end{eqnarray}
with
\begin{eqnarray}
	A_{\kappa}=\frac{\eta_+^{\kappa}-\eta_-^{\kappa}}{\sqrt{E^2-4J_LJ_R}}, ~\eta_{\pm}=\frac{E\pm \sqrt{E^2-4J_LJ_R}}{2},
\end{eqnarray}
{where, in general, the square root in the complex field has two branches. Hereafter, we use $\sqrt{\cdot}$ to represent one of the branches, and $-\sqrt{\cdot}$ to represent the other branch.}

Likewise, the relation can also be obtained by considering the backward transfer matrices connecting the $m$th with the $(m-1)$th non-mosaic sites, and thus the amplitude at mosaic site $\psi_{\kappa m-1}$ can be expressed as
\begin{eqnarray}
\psi _{\kappa m-1}=\frac{J_{R}^{\kappa -1}}{A_\kappa}\psi _{\kappa (m-1)}+\frac{J_LA_{\kappa-1}}{A_{\kappa}}\psi _{\kappa m}.
\label{eq:wave function1}
\end{eqnarray}
The details of derivation for the coefficients can be referred to in Appendix \ref{asec:transfer matrix}.

By substituting Eqs. \eqref{eq:wave function2} and \eqref{eq:wave function1} into the first line of Eq. \eqref{eq:recursion}, one can have a system of equations only involving non-mosaic sites, yielding
\begin{eqnarray}
J_R^\kappa\psi _{\kappa (m-1)}+J_L^\kappa\psi _{\kappa (m+1)}+\lambda A_{\kappa}\Delta_{\kappa m}\psi _{\kappa m}=B_{\kappa}\psi _{\kappa m}, \notag\\
\label{eq:scaled model}
\end{eqnarray}
with
\begin{eqnarray}
B_{\kappa}=EA_\kappa -2J_LJ_RA_{\kappa-1}.
\label{eq:En Trans}
\end{eqnarray}
If one regards $m$ as the { effective} site index and $\kappa$ just as a parameter, Eq. \eqref{eq:scaled model} is nothing but effectively represents a static Schr\"{o}dinger equation for a non-mosaic tight-binding model with $J_{L,R}^\kappa$ being the nonreciprocal hopping, $\lambda A_{\kappa}\Delta_{\kappa m}$ being on-site potential, and $B_\kappa$ being the eigenenergy. For convenience,  in the following, we will refer to the model (\ref{eq:scaled model}) as {\it the scaled model}, in contrast to the original model (\ref{eq:recursion}).

Comparing the scaled model with the non-mosaic model \eqref{eq:recursion} with $\kappa=1$, we have the following general mapping:
\begin{eqnarray}
	J_{L,R} \rightarrow J_{L,R}^\kappa,\,\lambda\rightarrow \lambda A_\kappa f_\kappa,\,E\rightarrow B_\kappa,\,\psi_{j}\rightarrow\psi_{\kappa m},
	\label{eq:mapping}
\end{eqnarray}
where the introduction of $f_\kappa$ reflects the relation between the potential functions $\Delta_{\kappa m}$ and $\Delta_m$ in the two models.

This general mapping \eqref{eq:mapping} reveals that if a wave function $\psi_j$ with an eigenenergy $E$ is localized at site $j$ under the potential function $\Delta_m$ in the non-mosaic model, then the localization also occurs at the {effective} site $m$ under the potential function $\Delta_{\kappa m}$ in the scaled model, described by the wave function $\psi_{\kappa m}$ with the eigenenergy $B_\kappa$. 
Moreover, it implies that the wave function is localized at the non-mosaic site $\kappa m$ in the mosaic model with finite $\kappa$. 
Therefore, given the critical point of localization determined by a function $f(J_L, J_R, \lambda,E)=0$ in the non-mosaic model, the critical point in the mosaic model can be determined by
\begin{eqnarray}
	f(J_L^\kappa, J_R^\kappa,\lambda A_\kappa f_\kappa,B_\kappa)=0,
	\label{eq:mosaic_critical}
\end{eqnarray}
through the general mapping \eqref{eq:mapping}.
Generally, the critical point of the mosaic model depends on $E$, which means that there exist mobility edges, even if there are no mobility edges for the non-mosaic model. This is a result of $A_\kappa$ being a function of $E$. 
In principle, if one knows the critical point of a {non-Hermitian} non-mosaic model, the critical point of the corresponding mosaic model can also be obtained using Eq. \eqref{eq:mosaic_critical}.

Furthermore, LE $\gamma(J_L,J_R,\lambda,E)>0$, which characterizes the decaying behavior of a localized state, has also been established for certain non-mosaic models with random or quasiperiodic disorders. In these models, the localized state at site $j_0$ takes the form $|\psi_j|\propto e^{-\gamma |j-j_0|}$.
Using this expression and the general mapping, one can also obtain the LE $\gamma(J_L^\kappa,J_R^\kappa,\lambda A_\kappa f_\kappa,B_\kappa)>0$ for the scaled model \eqref{eq:scaled model} with respect to the ``sites'' labeled by $m$. This results in a localized state given by
\begin{eqnarray}
|\psi_{\kappa m}|\propto e^{-\gamma (m-m_0)}=e^{-\frac{\gamma}{\kappa} (\kappa m-\kappa m_0)}	\label{eq:mosaic_LE}
\end{eqnarray}
at the non-mosaic site $\kappa m_0$, where $\gamma/\kappa$ represents the inverse of localization length or the LE \cite{CM} of the corresponding mosaic models with respect to the non-mosaic sites $\kappa m$.
It is worth noting that the critical point can also be obtained from the condition $\gamma(J_L^\kappa,J_R^\kappa,\lambda A_\kappa f_\kappa,B_\kappa)=0$, which is equivalent to Eq. \eqref{eq:mosaic_critical}.

{In principle, the general mapping (\ref{eq:mapping}) is applicable to any non-Hermitian mosaic models.} In the following, we will focus on a more specific case where the hopping can be parameterized as $J_L=te^{-g}$ and $J_{R}=te^{g}~(t,g\in \mathbb{R})$. Consequently, the scaled model \eqref{eq:scaled model} becomes
\begin{eqnarray}
te^{\kappa g}\psi _{\kappa (m-1)}+te^{-\kappa g}\psi _{\kappa (m+1)}+\lambda a_{\kappa}\Delta _{\kappa m}\psi _{\kappa m}=\varepsilon_{\kappa}\psi _{\kappa m}, \notag\\
\label{eq:nonrep-model}
\end{eqnarray}
as a {non-Hermitian} generalization of the Hermitian case discussed in Ref. \cite{Liu2022}.
Here, we replace $A_\kappa$ and $B_\kappa$ in Eq. (\ref{eq:scaled model}) with the following quantities: 
\begin{eqnarray}
	a_{\kappa}&=&\frac{1}{\sqrt{E^2/t^2-4}}\Bigg[\bigg(\frac{E/t+ \sqrt{E^2/t^2-4}}{2}\bigg)^\kappa\notag\\
	&&~~~~~~~~~~~~~~~~~~ -\left(\frac{E/t- \sqrt{E^2/t^2-4}}{2}\right)^\kappa\Bigg], \notag\\
	\varepsilon_{\kappa}&=&Ea_\kappa -2ta_{\kappa-1},
	\label{eq:ae}
\end{eqnarray}
 where $a_\kappa$ is dimensionless, $\varepsilon_\kappa$ has the dimension of energy, and we define $a_0=0$ for the correct expression of $\varepsilon_1$. 
 Specifically,
\begin{eqnarray}
	a_{1}&=&1,~~~~~~~~~~~~\varepsilon_1=E,  \notag\\
	a_{2}&=&E/t,~~~~~~~~~\varepsilon_2=E^2/t-2t,   \notag\\
	a_{3}&=&E^2/t^2-1,~\varepsilon_3=E^3/t^2-3E,
\end{eqnarray}
which will be used in the following applications. 
In terms of the parameters $\{t,g\}\in \mathbb{R}$, the general mapping \eqref{eq:mapping} becomes
\begin{eqnarray}
		t\rightarrow t,\,g \rightarrow \kappa g ,\,\lambda \rightarrow \lambda a_\kappa f_\kappa,\, E\rightarrow \varepsilon_\kappa,\,\psi_{j}\rightarrow\psi_{\kappa m}
		\label{eq:g-map}.
\end{eqnarray}
Apparently, when $g=0$, this mapping can be reduced to the Hermitian cases studied in Ref. \cite{Liu2022}.

To demonstrate the validity, we will apply the mapping \eqref{eq:g-map} to two specific {non-Hermitian} mosaic models in the following. For convenience, we will set $t=1$ as the energy unit and focus on the cases with $g>0$, which corresponds to a right-biased hopping.

\section{Applications}\label{sec:app}

\subsection{Non-Hermitian Aubry-Andr\'{e}-like mosaic model}

\begin{figure}[tbh] 
	\includegraphics[width=1\linewidth]{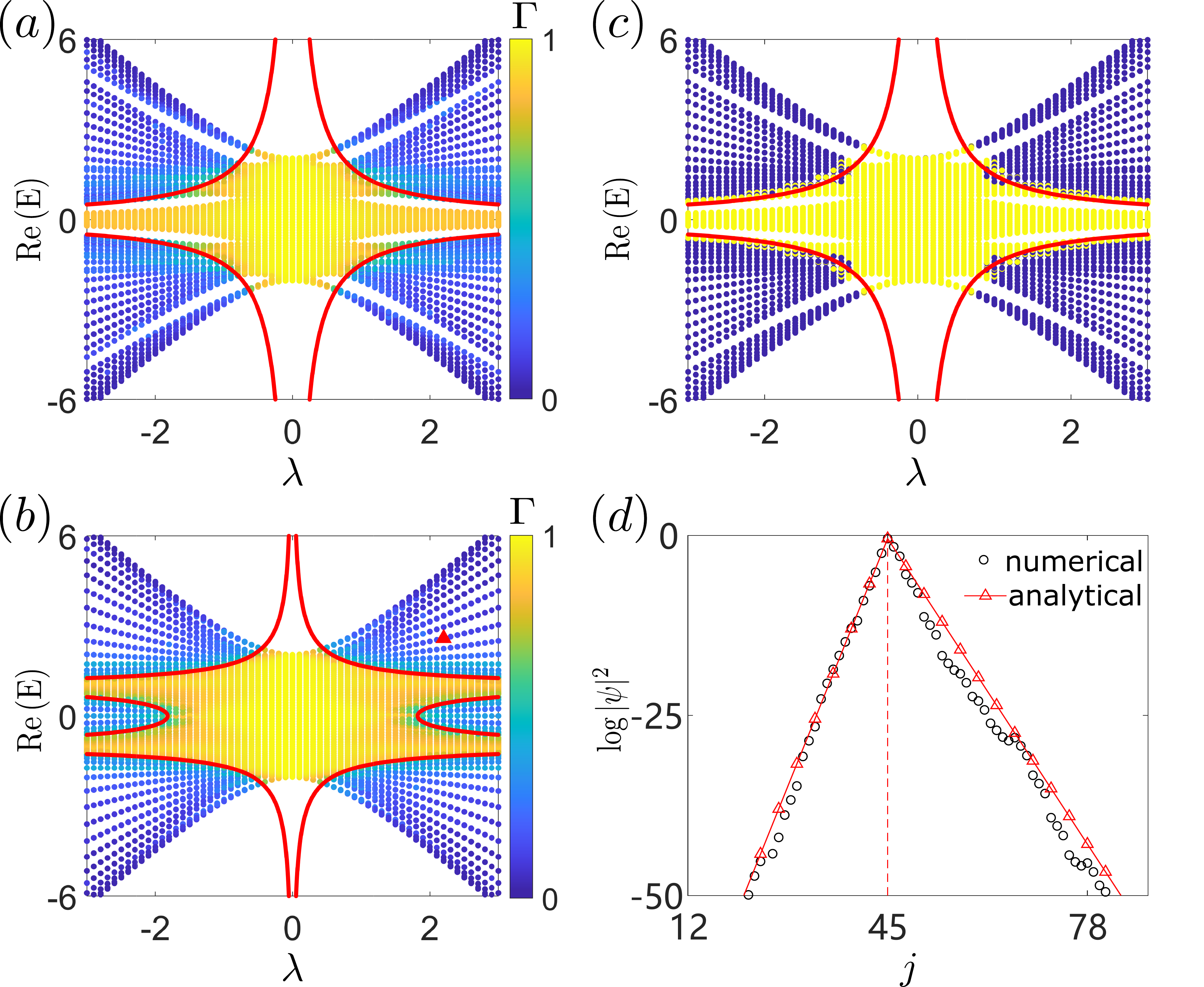}
	\caption{The fractal dimension $\Gamma$ as a function of the real part of the eigenenergy Re$(E)$ and a real potential strength $\lambda$, numerically calculated for (a) $\kappa =2$ and (b) $\kappa =3$ under PBCs with other parameters $g=0.2$, $\phi=0$, $L=F_{12}=144$, and $\beta=F_{11}/F_{12}=89/144$.		
	(c) The existence or not of the imaginary part of energy with the same parameters as in (a), where (non)zeros of Im$(E)$ with accuracy $10^{-6}$ are colored in blue (yellow). 
	The red solid lines in (a-c) are the mobility edges calculated analytically by Eq. \eqref{eq:MEAA}.
	(d) Comparison of a localized state marked by the red triangle ($E=2.5881,\lambda=2.2$) in (b) with the corresponding analytical form Eq. \eqref{eq:nonGAA wave function}, showing that the two LEs can well characterize the asymmetric localization in the {non-Hermitian} AA-like mosaic model. The dotted line indicates the center site of the localization.}
	\label{fig1}
\end{figure}

As the first application, we take a mosaic version of an AA-like model with nonreciprocal hopping and a complex potential, dubbed {\it {non-Hermitian} AA-like mosaic model} for the convenience of discussion in the following, which is described by Hamiltonian \eqref{eq:Ham} with the potential function in Eq. \eqref{eq:Disord} being
\begin{eqnarray}
	\Delta_j=\Delta_{\kappa m}=2\cos \left( 2\pi \beta \kappa m +\theta+i\phi\right),
	\label{eq:AA_pot}
\end{eqnarray}
where $\beta$ is an arbitrary irrational number and $\{\theta,\phi\}\in\mathbb{R}$ represent a complex phase shift.
The corresponding non-Hermitian AA-like non-mosaic model undergoes an Anderson localization at 
\begin{eqnarray}
|\lambda|=e^{g-|\phi|}.
\label{eq:cp-AA}
\end{eqnarray}
The detailed proof can be referred to in Appendix \ref{asec:proof}.
Notably, this critical point of localization is independent of the eigenenergy $E$; in other words, there are no mobility edges in the non-Hermitian AA-like non-mosaic model.
 
Based on the critical point Eq. (\ref{eq:cp-AA}), we now apply the mapping \eqref{eq:g-map} to study its mosaic counterpart. 
Since the difference between two potential functions $\Delta_{\kappa m}=2\cos (2\pi\beta\kappa m+\theta+i\phi)$ and $\Delta_{m}=2\cos (2\pi\beta m+\theta+i\phi)$ is only between the irrational numbers $\kappa\beta$ and $\beta$, the $f_\kappa$ in the mapping can be taken as $1$, because the critical points of Anderson localization for quasiperiodic models are independent of the values of irrational numbers \cite{Jitomirskaya1999,AvilaZhou2017}, although the eigenstates and eigenvalues under these two potentials are irrelevant in details. Thus, by making the replacement, $\lambda\rightarrow\lambda a_\kappa$ and $g\rightarrow\kappa g$, in Eq. (\ref{eq:cp-AA}), one can find that the mobility edges,
\begin{equation}
	|\lambda_\kappa(E)| = e^{\kappa g-|\phi|}/|a_{\kappa}|,
	\label{eq:AAH}
\end{equation}
emerge in the non-Hermitian AA-like mosaic model.
This result is identical to that of Ref. \cite{LiuChen2021c} obtained by Avila's global theory \cite{Avila2015}, verifying the correctness of the mapping.

Furthermore, one can also obtain the LEs in the non-Hermitian AA-like mosaic model by the mapping. 
From Ref. \cite{JiangChen2019}, we know that two LEs $\gamma\pm g>0$ can be used to characterize an asymmetrically localized state in a non-Hermitian model with nonreciprocal hopping, and the smaller one $\gamma-g$ determines the critical point of Anderson localization, where 
\begin{eqnarray}
	\gamma =\ln|\lambda|+|\phi|
\end{eqnarray}
is the LE for the corresponding symmetrically localized state in the non-Hermitian AA-like non-mosaic model with reciprocal hopping.

Therefore, with the same replacement, one can get the two LEs of the scaled model \eqref{eq:scaled model} as
\begin{eqnarray}
	\gamma^{(\pm)}_\kappa(E) =\ln|\lambda a_\kappa| + |\phi| \pm \kappa g,
	\label{eq:LE}
\end{eqnarray}
which depend on the eigenenergy $E$, leading to an asymmetrically localized state,
\begin{eqnarray}
	|\psi _{\kappa m}|\propto 
	\left\{ \begin{matrix}
		e^{-\frac{\gamma_\kappa^{(-)}}{\kappa}\left(\kappa m-\kappa m_0 \right)},&		m>m_0\\
e^{-\frac{\gamma_\kappa^{(+)}}{\kappa}\left(\kappa m_0-\kappa m\right)},&		m<m_0\\
	\end{matrix} \right.,
	\label{eq:nonGAA wave function}
\end{eqnarray}
with the peak being located at the non-mosaic site $\kappa m_0$. 
The right-biased asymmetry [i.e., $\gamma^{(-)}<\gamma^{(+)}$] of the localized state with respect to the center site $\kappa m_0$ results from the right-biased hopping (i.e., $g>0$) we set in advance.
This can be partially checked by setting $\gamma^{(-)}_\kappa(E)=0$, which just gives rise to the critical point in Eq. \eqref{eq:AAH}. 

For further verification, we compare the numerical results with the analytical ones for the cases of $\kappa=2$ and $3$ in Fig. \ref{fig1}, where, from Eqs. \eqref{eq:AAH} and \eqref{eq:LE}, the critical points and the LEs are specified as
{
\begin{eqnarray}
	|\lambda_{2}(E)|&=&\frac{e^{2g-|\phi|}}{|E|},~\frac{\gamma^{(\pm)}_2(E)}{2}=\frac{\ln|\lambda E|+|\phi|}{2}\pm g;  \notag\\
	|\lambda_{3}(E)|&=&\frac{e^{3g-|\phi|}}{|E^2-1|},~\frac{\gamma^{(\pm)}_3(E)}{3}=\frac{\ln|\lambda (E^2-1)|+|\phi|}{3}\pm g. \notag\\
	\label{eq:MEAA}
\end{eqnarray}
}
For the numerical calculation, we take $\beta=F_{s-1}/F_s$ as a rational approximation of an irrational number $\beta =\lim_{s\rightarrow\infty}F_{s-1}/F_{s}=(\sqrt{5}-1)/2$, and the system size $L=F_s$, where $F_s~(s=1,2,\cdots)$ is the $s$th Fibonacci number defined as $F_s=F_{s-1}+F_{s-2}$ with the first two numbers $F_1=F_2=1$. 
The fractal dimension \cite{EversMirlin2008},
\begin{eqnarray}
	\Gamma&=&-\ln I/\ln L,
\end{eqnarray}
can be used to characterize the localization of a state $\psi_j$ more clearly for a finite system than the commonly used inverse participation number (IPR) \cite{EversMirlin2008}, 
\begin{eqnarray}
	I&=&\sum_{j=1}^L|\psi_j|^4\Big/\Big(\sum_{j=1}^L|\psi_j|^2\Big)^2.
\end{eqnarray}

To demonstrate the transition of Anderson localization, of which the critical points are the same under {periodic boundary conditions (PBCs) and open boundary conditions (OBCs)} due to the blindness about boundary conditions for a bulk-localized state \cite{JiangChen2019}, we use PBCs for numerical calculations in Fig. \ref{fig1} to avoid non-Hermitian skin effect under OBCs \cite{YaoWang2018}, where all bulk states are aggregated together to one boundary (here the right boundary for our settings), leading to similar values of fractal dimension as bulk-localized states. 
Meanwhile, according to the bulk-bulk correspondence of a non-Hermitian model with nonreciprocal hopping when $\phi=0$, i.e., the critical point of Anderson localization under OBCs is also a cut between complex and real spectra under PBCs \cite{JiangChen2019} for Figs. \ref{fig1}(a) and \ref{fig1}(c), one can assume the reality of $E$ and abandon the complex-$E$ solutions during the numerical calculation of critical points. 
Due to Im$[(E(\lambda)]=0$, the mobility edges $E(\lambda)$ are just functions of curves, not surfaces in the Re$(E)$-Im$(E)$-$\lambda$ space.
The maximal number $2(\kappa-1)$ of mobility edges, as shown in Figs. \ref{fig1}(a) and \ref{fig1}(b), can also be expected from Eq. \eqref{eq:AAH}, where the highest power in $E$ can be $(\kappa-1)$.
The two LEs are demonstrated in Fig. \ref{fig1}(d) as two slopes from the peak site to the two shoulders. Due to the existence of side peaks on two shoulders beside the main peak, the numerical result can shift occasionally relative to the analytical result for some regions, but the preserved slope still reflects the correctness of the LEs derived from Eq. (\ref{eq:MEAA}).

\subsection{Nonreciprocal Ganeshan-Pixley-Das Sarma mosaic model}

\begin{figure}[tb]
	\includegraphics[width=1\linewidth]{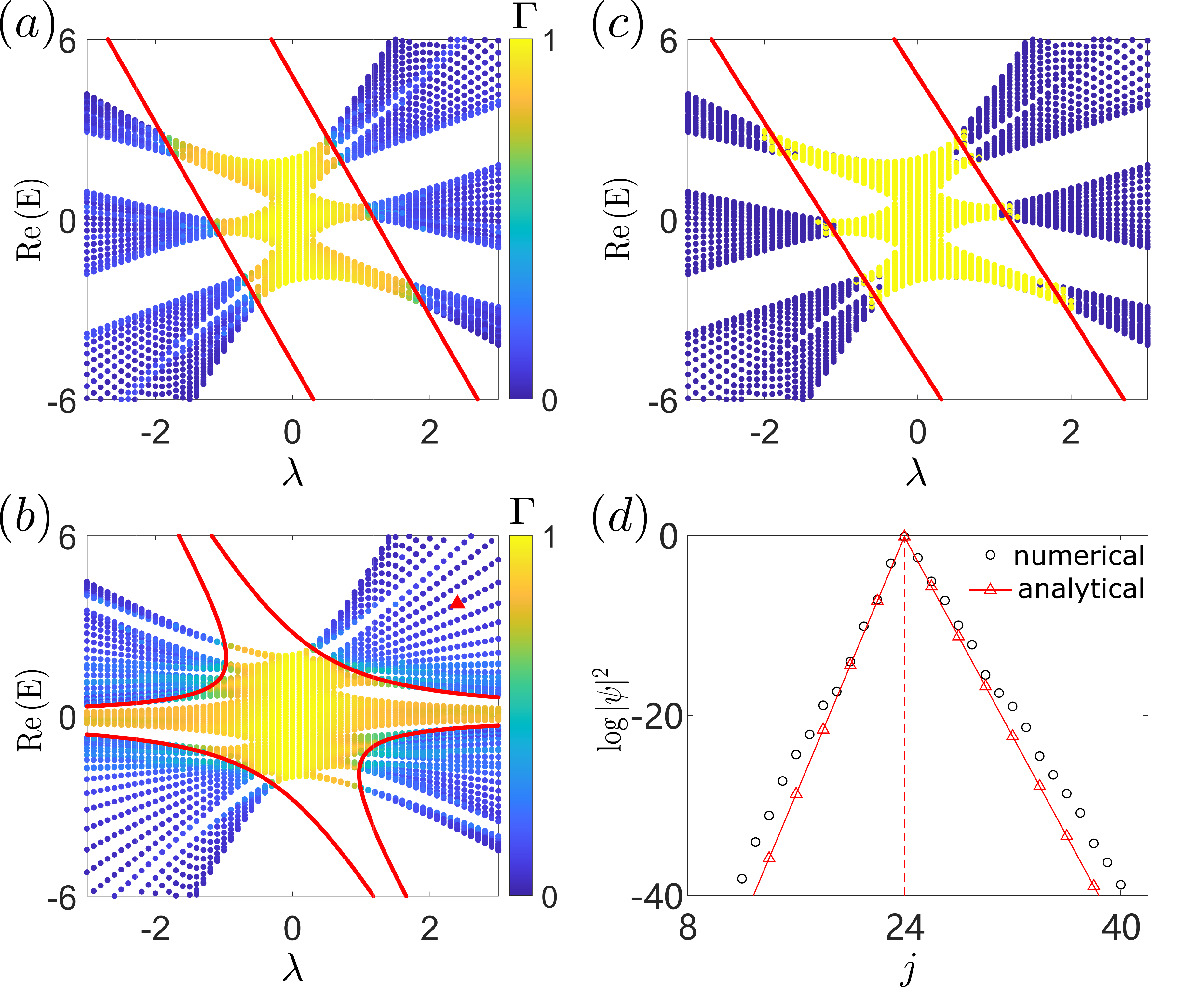}
	\caption{The same meanings and settings as in Fig. \ref{fig1}, except for (a,c) $\kappa=1$ and (b,d) $\kappa=2$. The mobility edges [red solid lines in (a-c)] are calculated analytically by Eq. \eqref{eq:GPDg}.
	The localized state marked by the red triangle in (b) is selected as $E=3.7328,\lambda=2.4$.
The parameter $b=0.5$ is chosen for all figures.}
	\label{fig2}
\end{figure}

The mobility edges in the non-Hermitian AA-like mosaic model emerge from a constant critical point of the non-mosaic counterpart. 
As another application, we consider the mosaic version of the GPD model {(with real potential strenghth $\lambda$)} \cite{GaneshanDasSarma2015} that incorporates nonreciprocal hopping, dubbed {\it the nonreciprocal GPD mosaic model}. Its non-mosaic counterpart already exhibits mobility edges. 
The Hamiltonian is described by Eq. \eqref{eq:Ham} with the following real potential function:
\begin{eqnarray}
\Delta_j=\frac{2\cos\mathrm{(}2\pi \beta j)}{1-b\cos\mathrm{(}2\pi \beta j)},
\end{eqnarray}
with the same definition of $\beta$ as the non-Hermitian AA-like model and the deformation parameter $b\in(-1,1)$.

For the non-mosaic case with nonreciprocal hopping, i.e., $\kappa=1$, the mobility edges \cite{LiuChen2021a}
\begin{eqnarray}
	\lambda(E) =\frac{1}{2}&\big[-bE\pm e^g\left(1+\sqrt{1-b^2}\right)&\notag\\
	&\pm e^{-g}\left(1-\sqrt{1-b^2}\right)\big],&
	\label{eq:GPDMEg}
\end{eqnarray}
can be obtained by Avila's global theory \cite{Avila2015} and the similarity transformation under OBCs \cite{JiangChen2019}. 
{Note here $\lambda(E)$ keeps real due to the reality of $E$ in this model.}
When $g=0$, it reduces to the reciprocal case with mobility edges being
\begin{eqnarray}
	\lambda(E)=\pm1-\frac{bE}{2},
	\label{eq:GPD}
\end{eqnarray}
which is first obtained by a generalized duality transformation \cite{GaneshanDasSarma2015},
and the LE
\begin{eqnarray}
	\gamma (E)=\ln \bigg| \frac{|bE+2\lambda |+\sqrt{(bE+2\lambda )^2-4b^2}}{2\left(1+\sqrt{1-b^2} \right)} \bigg|
	\label{eq:GPDLE}
\end{eqnarray}
is analytically obtained in Refs. \cite{LiuChen2021a} and \cite{WangLiu2021} with the aid of the Avila's global theory. The condition $\gamma(E)=0$ can also generate the critical point Eq. \eqref{eq:GPD}. 

With the aid of the similarity transformation under OBCs in Ref. \cite{JiangChen2019}, we can readily derive the two LEs, $\gamma(E)\pm g$, in the nonreciprocal GPD non-mosaic model ($\kappa=1$ and $g\ne0$), and the validity of the mobility edges given by Eq. \eqref{eq:GPDMEg} can be confirmed by the condition $\gamma(E)-g=0$.

{In the noneciprocal GPD mosaic model, the eigenenergy $E$ must be a real number for a localized state due to the similarity transformation under OBCs to a Hermitian model \cite{JiangChen2019}.
Therefore, we can safely apply the mapping (\ref{eq:g-map}) in the real field, and the absolute and squared root operations in Eq. \eqref{eq:GPDLE} retain their definitions in the real field.}

Finally, by the mapping \eqref{eq:g-map}, we obtain the two LEs in the nonreciprocal GPD mosaic model as follows
\begin{eqnarray}
	\frac{\gamma^{(\pm)}_\kappa(E)}{\kappa}&=&\frac{1}{\kappa}\ln \bigg|\frac{|b\varepsilon_\kappa+2\lambda a_\kappa|+\sqrt{(b\varepsilon_\kappa+2\lambda a_\kappa)^2-4b^2}}{2\left(1+\sqrt{1-b^2} \right)}\bigg| \notag\\
	&&\pm  g,
\end{eqnarray}
{where $a_\kappa$ and $\varepsilon_\kappa$ defined in Eq. \eqref{eq:ae} are real numbers due to the reality of $E$ for localized states and the system parameters.}
The mobility edges can be determined by the condition $\gamma^{(-)}_\kappa(E)=0$, or directly obtained by the mapping from Eq. \eqref{eq:GPDMEg}, yielding
\begin{eqnarray}
	\lambda_\kappa(E) =\frac{1}{2a_\kappa}&\big[-b\varepsilon_{\kappa}\pm e^{\kappa g}\left(1+\sqrt{1-b^2}\right)&\notag\\
	&\pm e^{-\kappa g}\left(1-\sqrt{1-b^2}\right)\big].&
	\label{eq:GPDg}
\end{eqnarray}

Similarly, we consider the cases of $\kappa=1$ and $2$ for demonstration, as shown in Fig. \ref{fig2}, where we use the same setting of $\beta$ and PBCs as in the aforementioned non-Hermitian AA-like mosaic model.
The mobility edges are also curves that separate real spectra and complex ones, and the number of mobility edges can be maximally $2(\kappa-1)$, as expected.

\section{Conclusion and discussion}\label{sec:conclusion}

We establish a general mapping (\ref{eq:mapping}) from {non-Hermitian} mosaic models to their non-mosaic counterparts. By utilizing the analytical expressions of critical points of localization or even the LEs of localized states in the non-mosaic counterparts, one can derive the analytical expressions of mobility edges and the LEs for {non-Hermitian} mosaic models. 
As demonstrations, we take the non-Hermitian AA-like mosaic model and the nonreciprocal GPD mosaic model to test the mapping, successfully yielding analytical expressions for the mobility edges and LEs of these models.

The mapping is applicable not only to quasiperiodic models but also to randomly disordered models and non-disorder models that undergo localization transitions. 
However, in one dimension, critical points or LEs for {non-Hermitian} models are less established. Therefore, in this paper, we only focus on the applications of the mapping to the mosaic versions of quasiperiodic models.
For instance, in Ref. \cite{Liu2022} the mobility edges for the mosaic version of the Wannier-Stark model with reciprocal hopping are obtainable due to the availability of the critical point in the non-mosaic counterpart. However, the absence of the analytical expression of LE prevents us from obtaining the critical point or mobility edges of the mosaic models with {\it nonreciprocal} hopping using this mapping.

\appendix

\section{Derivation for the transfer matrix}\label{asec:transfer matrix}

For the forward transfer matrix $F$ of Eq. \eqref{eq:Transfer} in the main text, defined as
\begin{eqnarray}
	\left( \begin{matrix}
		\psi_{\kappa m+2}\\
		\psi_{\kappa m+1}\\
	\end{matrix} \right)
	&=&\left( 
	\begin{matrix}
		E/J_L&		-J_R/J_L\\
		1&		0\\
	\end{matrix} \right)
	\left( \begin{matrix}
		\psi_{\kappa m+1}\\
		\psi_{\kappa m}\\
	\end{matrix} \right) \notag\\
	&\equiv&F\left( 
	\begin{matrix}
		\psi_{\kappa m+1}\\
		\psi_{\kappa m}\\
	\end{matrix} \right),
\end{eqnarray}
we can diagonaize $F$ as
\begin{eqnarray}
	F&=&	\left(
	\begin{matrix}
		E/J_L &		-J_R/J_L\\
		1&		0\\
	\end{matrix} \right)=U\Lambda U^{-1},
\end{eqnarray}
where
\begin{eqnarray}
	\Lambda=\left(\begin{matrix}
	\eta_+/J_L&	0\\
	0&		\eta _-/J_L\\
\end{matrix} \right)
\end{eqnarray}
is the diagonal matrix with 
\begin{eqnarray}
	\eta _{\pm}&=&\frac{E\pm \sqrt{E^2-4J_LJ_R}}{2},
\end{eqnarray}
and
\begin{eqnarray}
	U=\left(\begin{matrix}
	\eta _+/J_L& \eta _-/J_L	\\
		1&	1\\
	\end{matrix} \right)
\end{eqnarray}
is a unitary matrix with its inverse being
\begin{eqnarray}
	U^{-1}=\left(\begin{matrix}
		J_L&	-\eta _-\\
		-J_L&		\eta _+\\
	\end{matrix} \right)\Big/\sqrt{E^2-4J_LJ_R}.
\end{eqnarray}
{In general, the square root in the complex field has two branches. Hereafter, we use $\sqrt{\cdot}$ to represent one of the branches, and $-\sqrt{\cdot}$ to represent the other branch.}

Then, one can get
\begin{eqnarray}
		F^{\kappa -1}&=&U\Lambda^{\kappa-1}U^{-1} \notag\\
		&=&\left(
		\begin{matrix}
			A_{\kappa}/J_{L}^{\kappa -1}&-J_RA_{\kappa-1}/J_{L}^{\kappa -1}\\
			A_{\kappa -1}/J_{L}^{\kappa -2}&	-J_RA_{\kappa -2}/J_{L}^{\kappa -2}\\
		\end{matrix} \right),
\end{eqnarray}
where
\begin{eqnarray}
	A_{\kappa}&=&\frac{\eta _{+}^{\kappa}-\eta _{-}^{\kappa}}{\sqrt{E^2-4J_LJ_R}}.
\end{eqnarray}
Therefore, we obtain the Eq. \eqref{eq:wave function2} in the main text:
\begin{eqnarray}
	\psi _{\kappa m+1}=\frac{J_{L}^{\kappa -1}}{A_\kappa}\psi _{\kappa (m+1)}+\frac{J_RA_{\kappa-1}}{A_{\kappa}}\psi _{\kappa m}.
\end{eqnarray}

For the backward transfer matrix $B$, defined as
\begin{eqnarray}
	\left( \begin{matrix}
		\psi_{\kappa m-2}\\
		\psi_{\kappa m-1}\\
	\end{matrix} \right)
	&=&\left( 
	\begin{matrix}
		E/J_R&		-J_L/J_R\\
		1&		0\\
	\end{matrix} \right)
	\left( \begin{matrix}
		\psi_{\kappa m-1}\\
		\psi_{\kappa m}\\
	\end{matrix} \right) \notag\\
	&\equiv&B\left( 
	\begin{matrix}
		\psi_{\kappa m-1}\\
		\psi_{\kappa m}\\
	\end{matrix} \right),
\end{eqnarray}
we have
\begin{eqnarray}
	\left( \begin{array}{c}
		\psi _{\kappa (m-1)}\\
		\psi _{\kappa (m-1)+1}\\
	\end{array} \right) =B^{\kappa -1}\left( \begin{array}{c}
		\psi _{\kappa m-1}\\
		\psi _{\kappa m}\\
	\end{array} \right).
\end{eqnarray}
Likewise, we can get
\begin{eqnarray}
	B^{\kappa -1}=\left( \begin{matrix}
		A_{\kappa}/J_{R}^{\kappa -1}&	-J_LA_{\kappa-1}/J_{R}^{\kappa -1}\\
		A_{\kappa -1}/J_{R}^{\kappa -2}&	-J_LA_{\kappa -2}/J_{R}^{\kappa -2}\\
	\end{matrix} \right),\notag\\
\end{eqnarray}
and thus the Eq. \eqref{eq:wave function1} in the main text
\begin{eqnarray}
	\psi _{\kappa m-1}=\frac{J_{R}^{\kappa -1}}{A_{\kappa}}\psi _{\kappa (m-1)}+\frac{J_LA_{\kappa -1}}{A_{\kappa}}\psi _{\kappa m}.
\end{eqnarray}

It is worth noting that the above derivation should assumes that 
\begin{eqnarray}
	E\ne\pm2\sqrt{J_RJ_L},
\end{eqnarray}
which corresponds the defectiveness of the transfer matrices $F$ and $B$. However, we can treat these cases as limiting cases from nondefective $F$ and $B$.

\section{Proof for the critical point of the AA-like model in the complex field}\label{asec:proof}
For the conventional AA model \cite{AubryAndre1980}, which corresponds to the parameter setting $J_R=J_L=t$ and $V_j=2\lambda\cos(2\pi\beta j+\theta)$ with $\beta$ being an arbitrary irrational number and $\{t,\lambda,\theta\}\in \mathbb{R}$ in Eq. (\ref{eq:Ham}), it is well known that the critical point occurs at $|\lambda/t|=1$.
However, we cannot directly apply this expression in the mapping (\ref{eq:g-map}) because the effective on-site potential $\lambda a_\kappa$ in the scaled model (\ref{eq:nonrep-model}) is generally complex.
Therefore, it is necessary to determine the critical point in the complex field for the conventional AA model with a complex strength $\lambda\in\mathbb{C}$ of the on-site quasiperiodic potential.

The eigenvalue equations for the AA-like model with a complex strength $\lambda\in\mathbb{C}$ and a complex phase shift $\theta+i\phi~(\theta,\phi\in\mathbb{R})$ of potential can be written as
\begin{eqnarray}
	E\psi _{j}=t(\psi _{j-1}+\psi _{j+1})+2\lambda \cos(2\pi\beta j+\theta+i\phi)\psi _{j},\notag\\
\end{eqnarray}
which can be reexpressed as
\begin{eqnarray}
	\begin{pmatrix}
		\psi_{j+1}\cr
		\psi_j
	\end{pmatrix} = T^{(j)}(E,\theta,\phi)
	\begin{pmatrix}
		\psi_{j}\cr
		\psi_{j-1}
	\end{pmatrix},
\end{eqnarray}
by the transfer matrix (we set $t=1$ as the unit of energy),
\begin{eqnarray}
	T^{(j)}(E,\theta,\phi)= 
	\begin{pmatrix}
		E-2\lambda \cos (2\pi \beta j +\theta+ i \phi) &  -1\cr
		1 & 0 
	\end{pmatrix}.\notag\\
\label{aeq:transfer_matrix}
\end{eqnarray}

The LE can be defined as follows \cite{Avila2015}:
\begin{eqnarray}
	\gamma(E,\phi) &\equiv&\lim_{n\rightarrow\infty}\frac{1}{2\pi n}\int_0^{2\pi}\ln ||T_n(E,\theta,\phi)||d\theta,
	\label{aeq:LE}
\end{eqnarray}
where 
\begin{eqnarray}
	T_n(E,\theta,\phi) &=&\prod_{j=1}^{n}T^{(j)}(E,\theta,\phi)
\end{eqnarray}
and the norm of $T_n(E,\theta,\phi)$ is defined as
\begin{eqnarray}
	||T_n(E,\theta,\phi)||=\max\{\sqrt{\chi_1},\sqrt{\chi_2}\},
\end{eqnarray}
with $\chi_{1,2}$ being the two nonnegative eigenvalues of $T^\dag_n(E,\theta,\phi)T_n(E,\theta,\phi)$.

To find the expression of $||T_n(E,\theta,\phi)||$, i.e., the expression of the maximum eigenvalue of $T^\dag_n(E,\theta,\phi)T_n(E,\theta,\phi)$, we resort to the large $\phi$ limit, i.e., $\phi \rightarrow +\infty$,  and the transfer matrix (\ref{aeq:transfer_matrix}) can be approximated as
\begin{equation}
	T^{(j)}(E,\theta,\phi)= e^{ \phi }\big[e^{-i(2\pi\beta j+\theta)}\left(
	\begin{array}{cc}
		-\lambda  &  0 \\
		0 & 0 \\
	\end{array}
	\right)+o(1)\big].
\end{equation} 
Thus, the maximum eigenvalue of $T^\dag_n(E,\theta,\phi)T_n(E,\theta,\phi)$ can be approximated as $|\lambda e^{\phi}|^{2n}$, and from Eq. (\ref{aeq:LE}) one can get the LE in the large $\phi$ limit:
\begin{eqnarray}
	\gamma(E,\phi) =\ln|\lambda|+\phi+o(1).
	\label{aeq:LE-large_phi}
\end{eqnarray}

According to Avila's global theory \cite{Avila2015}, $\gamma (E,\phi)$ defined by Eq. (\ref{aeq:LE}) is a convex, piecewise linear function of $\phi$ with the slope $\partial\gamma(E,\phi)/\partial\phi$ of each piece being an integer. 
The large-$\phi$ limit Eq. (\ref{aeq:LE-large_phi}) suggests that the slope for any $\phi\ge0$ is restricted to either $0$ or $1$. 
The discontinuity of the slope occurs when $E$ becomes an eigenenergy $E_s$ of the system except for $\gamma(E_s,\phi)=0$ \cite{Johnson-1986}, which represents the extended states.
This implies that the LE for an eigenenergy of the system can only be expressed as
\begin{eqnarray}
	\gamma(E_s,\phi)= 
		\max\{\ln| \lambda|+\phi, 0\}.
\end{eqnarray}
Considering that the LE is an even function of $\phi$ due to the property $T_n(E,\theta,\phi)\in SL(2,\mathbb{C})$ \cite{Avila2015}, the above formula can be further modified as
\begin{eqnarray}
	\gamma(E_s,\phi)= 
	\max\{\ln| \lambda|+|\phi|, 0\}.
\end{eqnarray}
Therefore, the critical point of localization is obtained by setting $\ln|\lambda|+|\phi|=0$, yielding
\begin{eqnarray}
	|\lambda|=e^{-|\phi|}.
\end{eqnarray}
Note that the LE depends only on the magnitude $|\lambda|$ of the complex potential strength, independent of its phase. 
This proof is similar to the scenario when $\lambda\in \mathbb{R}$ in Ref. \cite{LiuChen2021b}.

For the case extended to the nonreciprocal hopping (e.g., $J_L=te^{-g}$ and $J_R=te^{g}$ with $g> 0$), one can just use the similarity transformation as described in Ref. \cite{JiangChen2019} to obtain the two LEs,
\begin{eqnarray}
	\gamma^{(\pm)}(E,\phi)= \gamma(E,\phi)\pm g,
\end{eqnarray}
for the asymmetric localized states.

%
%

\begin{acknowledgments}
	This work was supported by the National Natural Science Foundation of China (Grant No.~12204406), the National Key Research and Development Program of China (Grant No.~2022YFA1405304), and the Guangdong Provincial Key Laboratory (Grant No.~2020B1212060066).
\end{acknowledgments}

\bibliographystyle{iopart-num}
\bibliography{ref}

\end{document}